\documentclass[ aps, onecolumn, prl, superscriptaddress, preprint]{revtex4-1}

\usepackage{algorithm2e, amsfonts, amsmath, amsbsy, appendix, bm, csquotes, dcolumn, dsfont, graphicx, mathrsfs, microtype, nameref, pslatex, soul, url, xcolor, xr, xr-hyper, hyperref}
\usepackage[T1]{fontenc}
\usepackage[margin=0.8in]{geometry}
\usepackage[sort&compress]{natbib}
\usepackage[skins,theorems]{tcolorbox}

\newcommand{\ee}[1]{\begin{align}#1\end{align}}

\newcommand{\bs}[1]{{\boldsymbol{#1}}}
\newcommand{\Cdot}{\! \cdot \!}

\bibstyle{ieeetr}

\begin{document}
\title{{Consensus between Epistemic Agents is Difficult}}
\author{Damian Rados\l aw Sowinski}
\email{Damian.Sowinski@Dartmouth.EDU}
\affiliation{Dartmouth College, Hanover NH 03755}
\author{Jonathan Carroll-Nellenback}
\email{Jonathan.Carroll@Rochester.EDU}
\affiliation{University of Rochester, Rochester NY}
\author{Jeremy DeSilva}
\affiliation{Dartmouth College, Hanover NH 03755}
\author{Adam Frank}
\affiliation{University of Rochester, Rochester NY 14627}
\author{Gourab Ghoshal}
\affiliation{University of Rochester, Rochester NY 14627}
\author{Marcelo Gleiser}
\affiliation{Dartmouth College, Hanover NH 03755}
\author{Hari Seldon}
\affiliation{Streeling University, Trantor}
\date{\today}
\begin{abstract}
{\fontsize{8}{11}}
\noindent
We introduce an epistemic information measure between two data streams, that we term {\it influence}. 
Closely related to transfer entropy, the measure must be estimated by epistemic agents with finite memory resources via sampling accessible data streams. 
We show that even under ideal conditions, epistemic agents using slightly different sampling strategies might not achieve consensus in their conclusions about which data stream is influencing which. 
As an illustration, we examine a real world data stream where different sampling strategies result in contradictory conclusions, explaining why some politically charged topics might exist due to purely epistemic reasons irrespective of the actual ontology of the world.  
\end{abstract}
\maketitle
\tableofcontents

\section{Introduction}
How do epistemic agents (EA)---rationally thinking beings with belief distributions over states of their world---judge information flows occurring throughout the complex systems they partake in?
How does their finiteness---in both space (access) and time (memory)---affect these judgements?
In all but the simplest of cases, information theoretic measures such as channel capacity, Shannon entropy, mutual information, transfer entropy, and others are defined with respect to empirically inaccessible joint probability distributions \cite{shannon1948mathematical, dembo1991information, cover1991information, schreiber2000measuring, lizier2008local, wibral2014directed, caticha2008lectures}.
An interdisciplinary problem, unbiased estimators are typically employed to mine finite datasets in an attempt to characterize these distributions\cite{wolpert1995estimating, agapiou2017importance, aguilera2020robust, hollingsworth2021efficient, rotskoff2021active}.
We demonstrate that groups of completely rational individuals with access to identical qualia to shape their beliefs, analogously formed by spatio-temporally limited sampling of sensory streams, may infer opposing judgements with regards to the relationships between those streams.

Cox's theorem sets probability theory on a firm epistemological foundation by establishing analytically the Bayesian interpretation of probability as {\it belief} 
\footnote{In this context, both $b\approx 1$ AND $b\approx 0$ are referred to as {\it strong beliefs}. A weak belief is typically the case $p\approx 1/2$ since we then refer to the space of possible experience with reference to a binary outcome - one that DID or DID NOT occur. For the more general case it is better to examine the entropy of belief relative to the maximum entropy.} 
\cite{cox1946probability, cox1963algebra, harms1998use, caticha2007information, jaynes2003probability}.
From an information theoretic perspective, belief is related to the uncertainty, or hidden information content, of experience \cite{jaynes1990probability, caticha2007information, harms1998use, caticha2008lectures}.
An epistemic concept, its connection to ontological thermodynamic entropy solved the problem of Maxwell's demon, and established a flourishing research program that has blurred the boundary between epistemology and ontology in an ongoing effort to unify information theory into physics\cite{szilard1929entropieverminderung, brillouin1953negentropy, doi:10.1119/1.1937760, landauer1961irreversibility, bennett1982thermodynamics}.
It is in this sense that we use the term {\it belief} in lieu of {\it probability}, and {\it information} instead of {\it entropy}.
Bayes' Rule is a roadmap for describing how epistemic agents (EA) learn about the world they find themselves embedded and interacting with  \cite{caticha2008lectures, caticha2006updating, ramsey2016truth}.
Each experience serves to update an EA's prior beliefs into posterior ones. 
The dynamics of belief emerges through the continuity of experience; an EA's judgements and actions being functionals of their dynamical belief \cite{caticha2002entropic, caticha2015entropic}.

Judgements about the world require EAs to store data of their observed past in some physical substrate (memory).
Any mass, $M$, playing the part of the substrate is capable of storing a maximal amount of information, known as the Bekenstein bound, $H_\text{max} \sim M^2$ \cite{PhysRevD.7.2333, prokopenko2014transfer}.
Meanwhile, the processing of information requires accessing memory at a rate ensuring fidelity, a procedure limited by the speed of light and uncertainty principle, the Bremmermen bound \cite{bremermann1982minimum, lloyd2000ultimate}, $\nu\sim M^{-1}$.
Finite computation time is a necessary characteristic of data-processing entities navigating an ever changing environment, implying a finite mass, which, in turn, implies a maximal memory.
An immediate corollary is that EAs are only capable of {\it sampling} history for use in their judgements, as a temporal continuum would require infinite memory.
In the simplest case, memory limitations mean there are two choices available to the EA's architecture: How often it takes a {\it snapshot} of the world, $\Delta t$, and how many of these {\it snapshots} it retains, $N$. 

This letter shows how the choice of architecture affects the conclusions available to an EA about its world by introducing a novel asymmetric measure of information flow called {\it influence}. 
As an illustration of our method, we examine a real world problem involving the interpretation of data concerning the CO$_2$ content and temperature of the atmosphere. 
Irrespective of the ontology of the world, epistemology drives policy, and we claim that our analysis reveals the etiology behind the ongoing {\it Weather versus Climate} argument.
Our primary result is that variance in the architectures of EAs makes disagreements inevitable, and consensus difficult.\\

\begin{figure*}[ht!]
    \centering
    \includegraphics[width=0.8\textwidth]{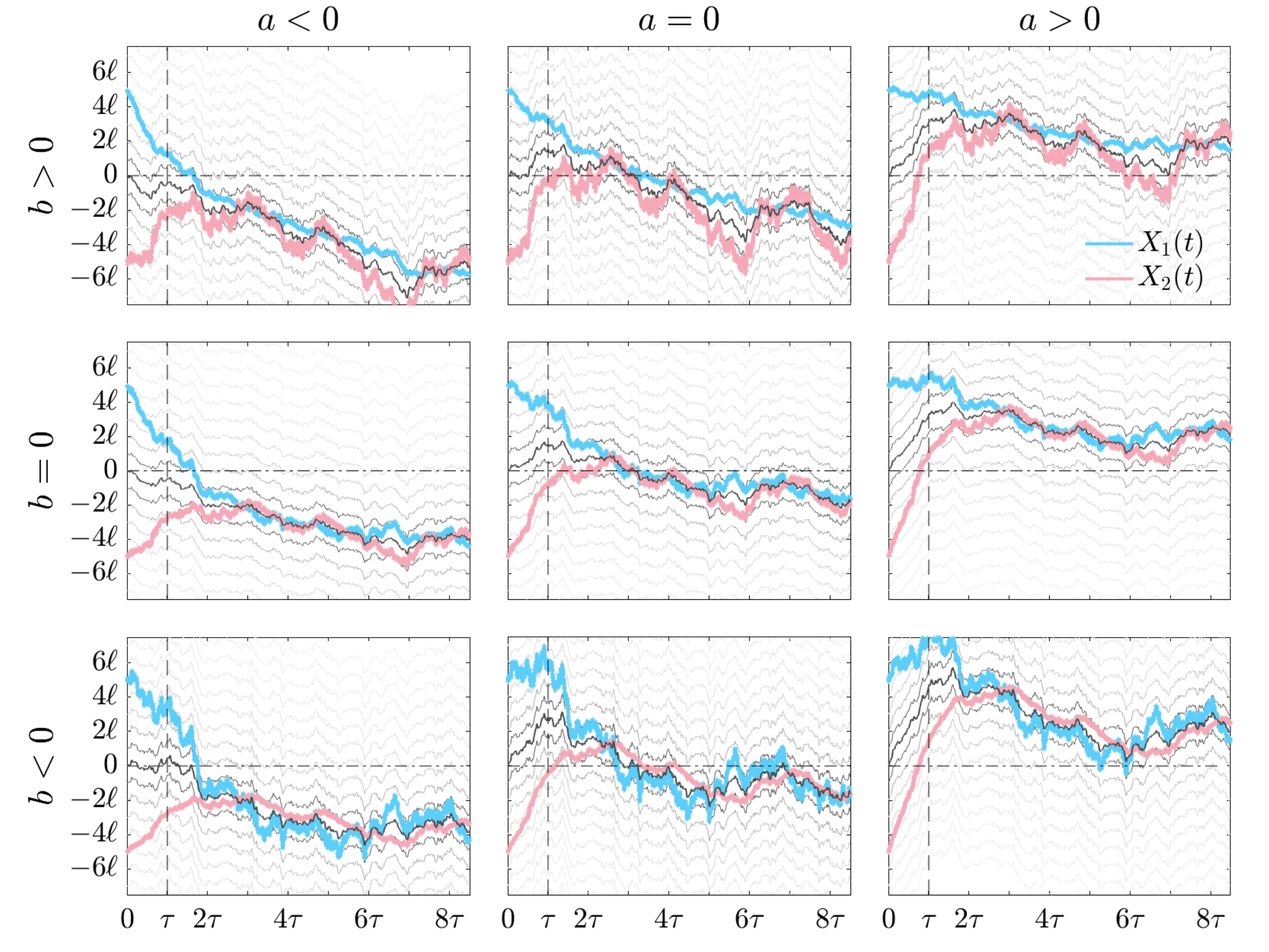}
    \caption{
    Instances of paths generated by equations of motion, Eq.[\ref{eq: X1 EoM}-\ref{eq: X2 EoM}], with the same seed for each pseudorandom generator used to simulate the heat bath.
    Axes have been scaled to the natural length and time scales, $\ell$ and $\tau$, respectively. 
    Paths are initialized 10$\ell$ units from each other. 
    Note the existence of transient behavior decaying with timescale $\tau$ followed by steady state behavior dominated by stochasticity. 
    The black line represents the mean of the two processes while the grey lines of decreasing opacity are an integer number of $\ell$s away from the mean. 
    }
    \label{fig: arxiv1}
\end{figure*}

\begin{figure*}[ht!]
    \centering
    \includegraphics[width=0.8\textwidth]{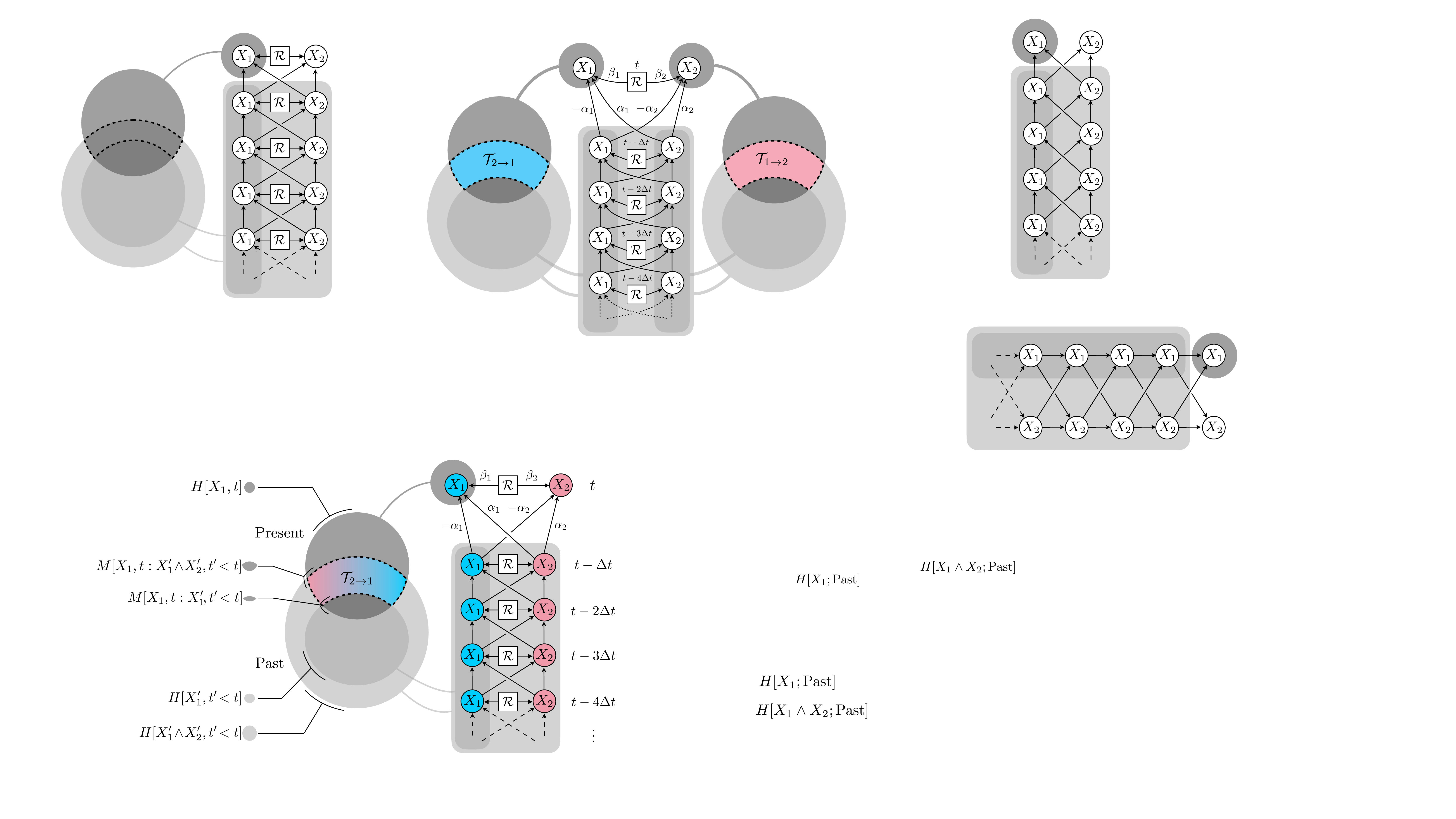}
    \caption{
    An information diagram of transfer entropy in our toy model. 
    To the right we have our coupled stochastic processes $X_1$ and $X_2$, as well as the heat reservoir, $\mathcal R$, to which they are connected. 
    Coupling constants are labelled $\alpha_i$ and $\beta_i$.  
    The {\it present} is at time $t$, and the temporal discretization scale is $\Delta t$. 
    To the left we have an information diagram where each circle represents the entropy of that quantity. 
    The present and past entropies are shown as bubbles, and the mutual information is the intersection of these bubbles. 
    The transfer entropy is labeled in relation to these mutual informations. 
    Note that this diagram has a mirror image on the other side of the processes (not shown) that represents the reverse $\mathcal T_{1\rightarrow 2}$ calculation.}
    \label{fig: arxiv2}
\end{figure*}

\section{Model}
First lets examine the relevant information measures necessary to model data gathering by an EA, and then build an analytically tractable toy model.
Let us focus on two processes, each generating a stream of qualia\cite{loar2003qualia}.
There is uncertainty associated with the future value of each of the qualia, and the processes could be correlated to one another above and beyond any sort of autocorrelation with their past.

Transfer entropy~\cite{massey1990causality, schreiber2000measuring} quantifies the extent to which past correlations between two processes reduce the present uncertainty in either process. 
That is, the extent to which an EA can predict the present of one of the processes, given past knowledge of both. 
For two processes, $X_1(t)\wedge X_2(t)$, it is defined as the excess information gained by knowing the past of process $2$ in addition to that of process $1$, 
\begin{align}
    \mathcal T_{2\rightarrow 1}\!=\!H[X_1,t|X_1',t'\!<\!t]\!-\!H[X_1,t|X_1'\wedge X_2',t'\!<\!t],
\end{align}
where $H[X]$ is the hidden information, i.e., the Shannon entropy of $X$, and $H[X|Y]$ is the conditional entropy of $X$ conditioned on $Y$ \cite{shannon1948mathematical}.
For analytical purposes ahead, it is useful to rewrite the transfer entropy in terms of the mutual information $M[X:Y]$ -  a measure of how much information is contained in the correlations between random variables $X$ and $Y$ - expressed as the Kullback-Leibler divergence between joint and product distributions, $M[X:Y]=\mathcal D_{KL}(\rho_{XY}||\rho_X\!\otimes\!\rho_Y)$\cite{kullback1951information}. 
Using this, we can write an alternative expression for the transfer entropy\cite{cover1991information, dembo1991information, gleiser2018we}:
\begin{align}\label{eq: MI form of transfer entropy}
    \mathcal T_{2\rightarrow 1}\!=\! M[X_1,t\!:\!X_1'\!\wedge\! X_2',t'\!\!<\!t]\!-\!M[X_1,t\!:\!X_1',t'\!\!<\!t]. 
\end{align}
For a finite EA, memory limitations preclude storage of the entire history, so a discrete subset of events is taken at $t'\in\{t-n\Delta t\}_{n=1,\dots, N}$, where $N$ is the length of the past stored alongside the present.
A schematic of the information diagram is shown in Fig \ref{fig: arxiv2}.

The processes in question can be treated symmetrically---how does knowing how $X_2$'s history decreases the uncertainty in $X_1$'s present compare to knowing how $X_1$'s history reduces the uncertainty in $X_2$'s present?
This treatment motivates our asymmetric measure of {\it influence},
\begin{align}
    \mathcal I_{ij}=\frac{\mathcal T_{i\rightarrow j}-\mathcal T_{j\rightarrow i}}{\mathcal T_{i\rightarrow j}+\mathcal T_{j\rightarrow i}}.
\end{align}
Influence is bounded on the interval $[-1,1]$, saturating iff the transfer entropy in only one direction vanishes.
This will occur when one of the processes is deterministic with vanishing stochasticity.
Since $\mathcal T_{i\rightarrow j}\ge 0$, we need only worry about the denominator when both directions vanish - we define $\mathcal I_{ij}\equiv0$ in this case. 
We note that this interpretation of {\it influence} is based entirely on correlations, and therefore should not be interpreted as a causal measure \cite{james2016information, PhysRevLett.103.238701,  lizier2010differentiating}.
This seems to be a recurring mistake in the literature, as transfer entropy reduces to the misleadingly named Granger-{\it Causality} in simple cases\cite{PhysRevLett.103.238701}.

\begin{figure*}[t!]
    \centering
    \includegraphics[width=\textwidth]{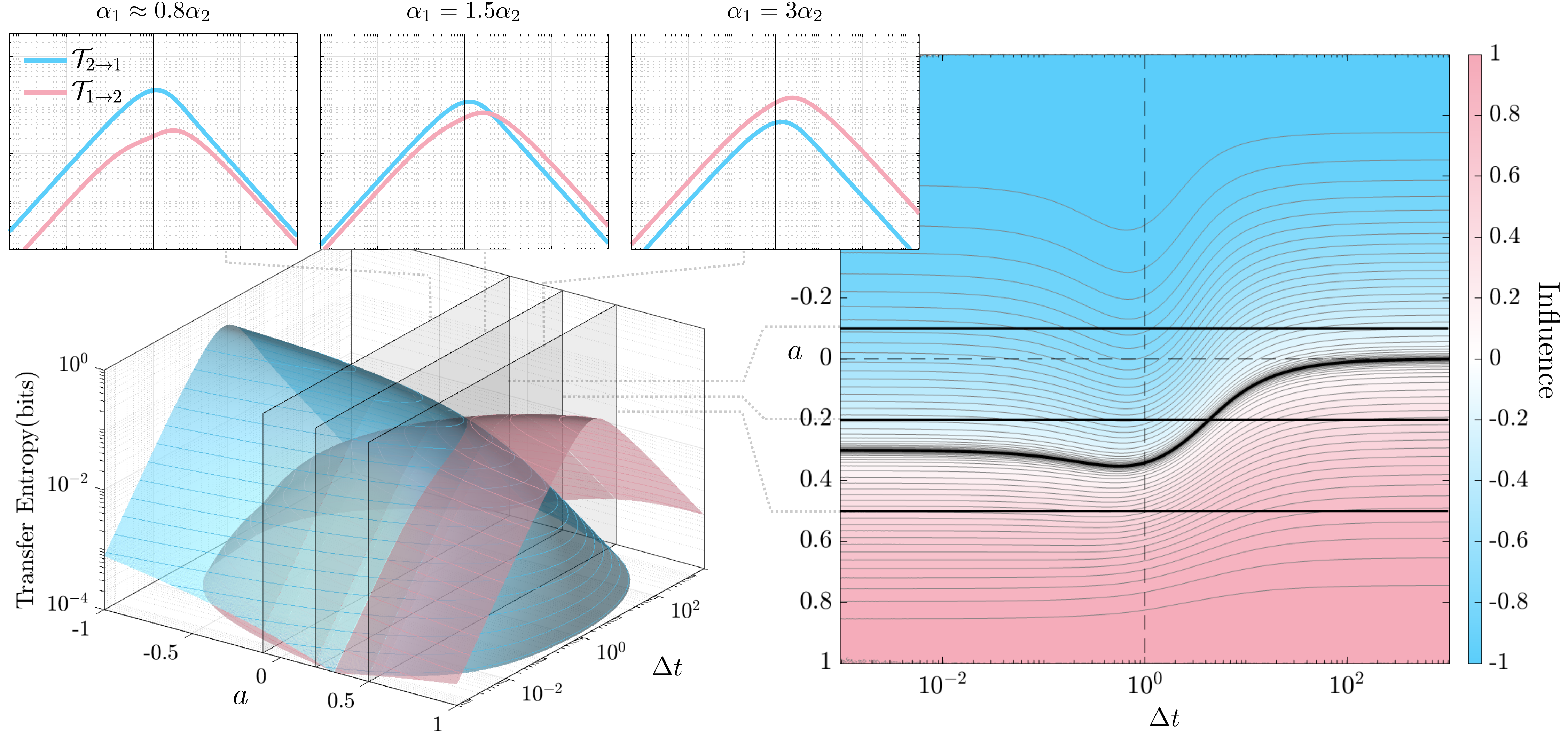}
    \caption{
    (A) The influence contours in our toy model across $a\in[-1,1]$ and $\Delta t\in[10^{-3},10^3]$ for $b=0.03$ and $N=4$.
    The three slices, $I,I\!I,$ and $I\!I\!I$, are taken at $a=-0.1,0.2,0.5$, respectively.
    (B) The transfer entropy surfaces from which (A) was constructed, with slices shown. 
    (C) The three slices are displayed. 
    Note the crossover for $a=0.2$. }
    \label{fig: arxiv3}
\end{figure*}

\begin{figure*}[t!]
    \centering
    \includegraphics[width=\textwidth]{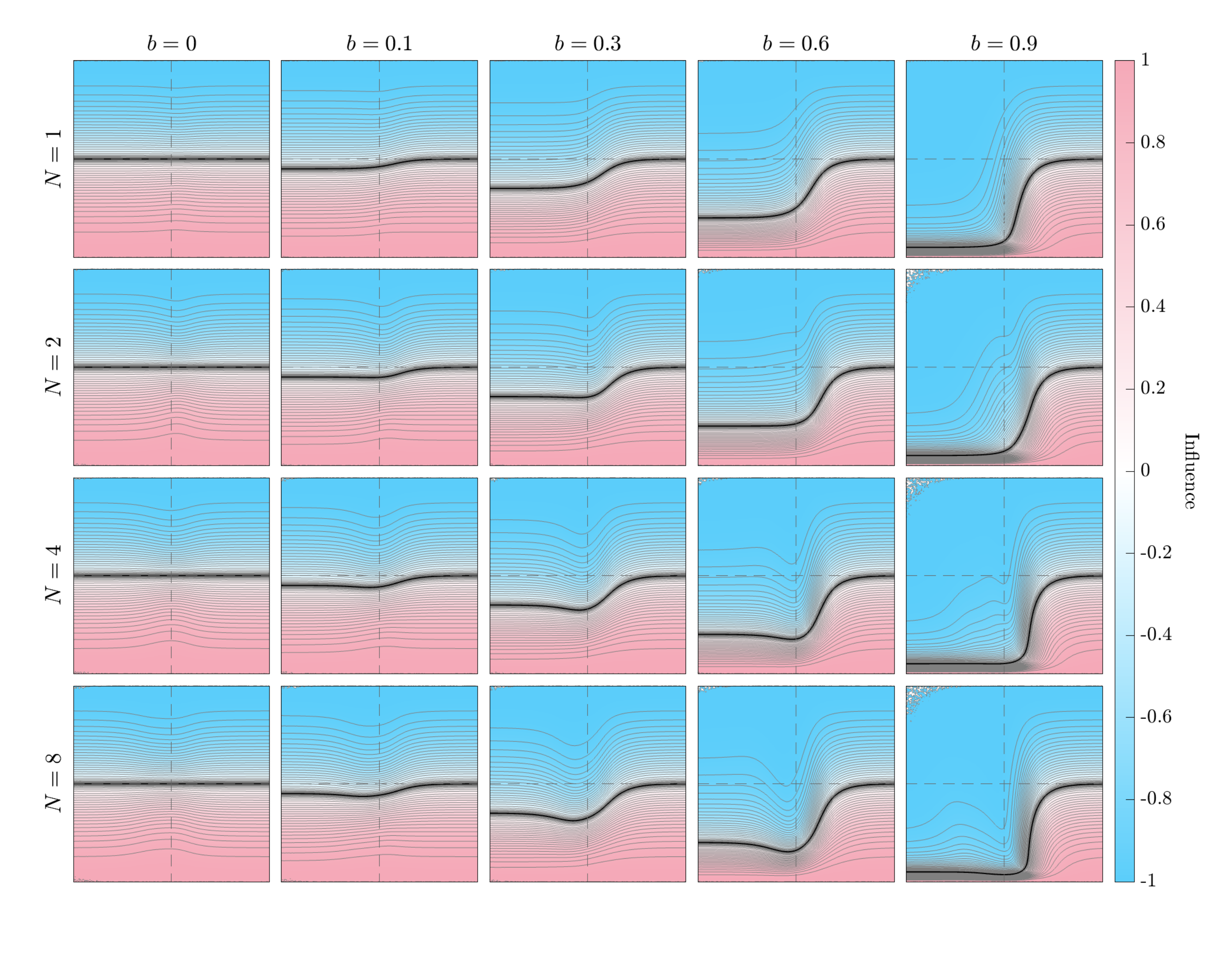}
    \caption{
    The Influence is plotted for other values of $b$ and $N$, with vanishing influence represented by the black line. 
    Each contour plot has the same axes as (A). 
    For negative values of $b$ the diagrams would have a flipped color scheme. 
    Note how for constant $a$ values the flipping of the sign of the influence is a generic feature across a large portion of the parameter space.}
    \label{fig: arxiv4}
\end{figure*}

We now construct an explicit toy model of coupled experiences to examine how a finite EA judges the influence between them.
Consider two processes modelled by a pair, $X_1(t)\wedge X_2(t)$, with evolution equations
\begin{align}\label{eq: X1 EoM}
\frac{d X_1}{dt}=-  \alpha_1(X_1-X_2)+\beta_1\eta_1\\
\label{eq: X2 EoM}
\frac{d X_2}{dt}=-  \alpha_2(X_2-X_1)+\beta_2\eta_2.
\end{align}
with initial conditions $X_1(t_0)\wedge X_2(t_0)$.
Here $\alpha_i,\beta_i\in\mathds{R}^+$ parameterize the deterministic and stochastic contributions, respectively, to the equations of motion.
The $\eta_i$ are independent white noise contributions that satisfy $\langle\eta_i(t)\rangle=0$ and $\langle \eta_i(t)\eta_j(t')\rangle = \delta_{ij}\delta(t-t')$.

The coupling constants $\alpha_i,\beta_i$ define natural time and length scales,
\begin{align}
\tau = \frac{1}{\alpha_1+\alpha_2}\hspace{10pt}\ell = \frac{\sqrt{\alpha_1+\alpha_2}}{\beta_1+\beta_2},
\end{align}
with which we dimensionalize all variables\cite{suppMaterial}. 
Fig.\ref{fig: arxiv1} shows several instances of the paths generated by Eqs. \ref{eq: X1 EoM} and \ref{eq: X2 EoM} scaled to these for several values of the coupling constants. 
The key point here is that this model has an explicit coupling which introduces correlations between the past and present of both processes. 
An EA can compute the information transfer and net intra-pair influence by sampling the processes at regular intervals, $\Delta t$, and storing up to $N$ samples given their memory capacity.
The quantities that determine the behavior of the processes' information dynamics are
\begin{align}
    a=\frac{\alpha_1-\alpha_2}{\alpha_1+\alpha_2}\hspace{10pt} b=\frac{\beta_1-\beta_2}{\beta_1+\beta_2},
\end{align}
which we call the deterministic and stochastic asymmetry parameters, respectively.

Eqs. \ref{eq: X1 EoM} and \ref{eq: X2 EoM} can be solved using projection operators\cite{suppMaterial}.
It is worth noting that the diagonalized system decomposes into a Wiener process for the center of mass motion, and an Ornstein-Uhlenbeck process for the separation. 
Both of these are Gaussian processes, which allows us to exactly solve for the analytical form of the influence as a function of $(\Delta t,N)$ of the pair using Eq. \ref{eq: MI form of transfer entropy}, though the resulting expression is quite unyieldy.
We go over the arduous procedure for doing so in the supplementary material\cite{suppMaterial}, and plot the results in Fig. \ref{fig: arxiv1}

The central contour plot of Fig. \ref{fig: arxiv3} shows the influence between two processes across the full domain $a\in[-1,1]$ as estimated by EAs with a range of memory resources $\Delta t\in[10^{-3},10^3]$. 
In the bottom-left surface plot (B) of Fig. \ref{fig: arxiv3}, we see the corresponding transfer entropy from process $1$ to process $2$ (blue), and vice-versa.
The dominating process is the one with the weaker deterministic coupling; the other process is pulled towards it, as seen in the left panel of Fig.\ref{fig: arxiv1}.
Three slices, labelled $I,I\!I,$ and $I\!I\!I$, are taken of these surfaces for constant values of the deterministic asymmetry parameter, namely $a\in\{-0.1,0.2,0.5\}$ and displayed in the upper-left three panels (C).
For the first (last) of these, the transfer entropy in the direction $2\rightarrow 1$ ($1\rightarrow2$) is always larger.  
Thus an EA would interpret the influence as always being unidirectional in both cases, irrespective of the temporal discretization $\Delta t$ employed. 
In the middle panel however there is a cross-over of transfer entropy at a particular value of temporal discretization.
The exact value of $\Delta t$ at which this occurs is not important in our discussion, as the far right plot of the influence shows that there is a wide range of deterministic asymmetry parameters that share this feature.
What is important is that the sign of the influence judgement made by the EA will depend on the temporal discretization. 
If the EA samples the processes at longer timescales, they will believe that process $2$ is influencing process $1$. 
But for a shorter sampling timescale, the EA will reach the opposite conclusion. 
Re-framing our analysis to multiple EAs that do not have an agreed upon sampling timescale, there exists the possibility that independent EAs will reach contradictory conclusions concerning how influential the two processes are amongst themselves. 
Our analysis indicates that consensus amongst epistemic agents is difficult.

The existence of contradictory conclusions for finite EAs is not simply dependent on the temporal discretization $\Delta t$ employed, but also the memory usage of the EAs captured by the parameter $N$.
The plot array in Figure \ref{fig: arxiv4} shows multiple instances of the central contour plot for varying values of the noise asymmetry parameter, $b$, and the memory usage of the EA, $N$. 
The black line in these panels shows where the influence flips, and the rows show the behavior of this feature as the memory usage $N$ of an EA increases. 
Since the line moves with increasing $N$, there are points of constant asymmetry parameters and temporal discretization that nonetheless lead to contradictory conclusions due to different memory capacities. 
Even if a consortium of EAs agrees upon a temporal discretization with which to sample the two processes, differences in memory usage once again introduce the possibility of contradictory conclusions. 
We note that the effect of $N$ appears weaker than that of $\Delta t$ as the locus of influence flips (the black line in these panels) changes slightly with $N$ and appears to saturate by $N \sim 8$.\\

\section{Climate Change vs Weather}\label{sec: 4}
We use climate data gathered at Mauna Loa Observatory - both CO$_2$ content and local temperatures - to explore an epistemological explanation of the weather-climate dispute\cite{dataMaunaLoa, dataMaunaLoa2, koutsoyiannis2020atmospheric}.
As most people do not have the priors of a scientist, let alone a climatologist or geophysicist, it is reasonable to model this argument as an EA with a limited history of a decade attempting to refine their beliefs.  

\begin{figure}[ht!]
    \includegraphics[width=0.6\columnwidth]{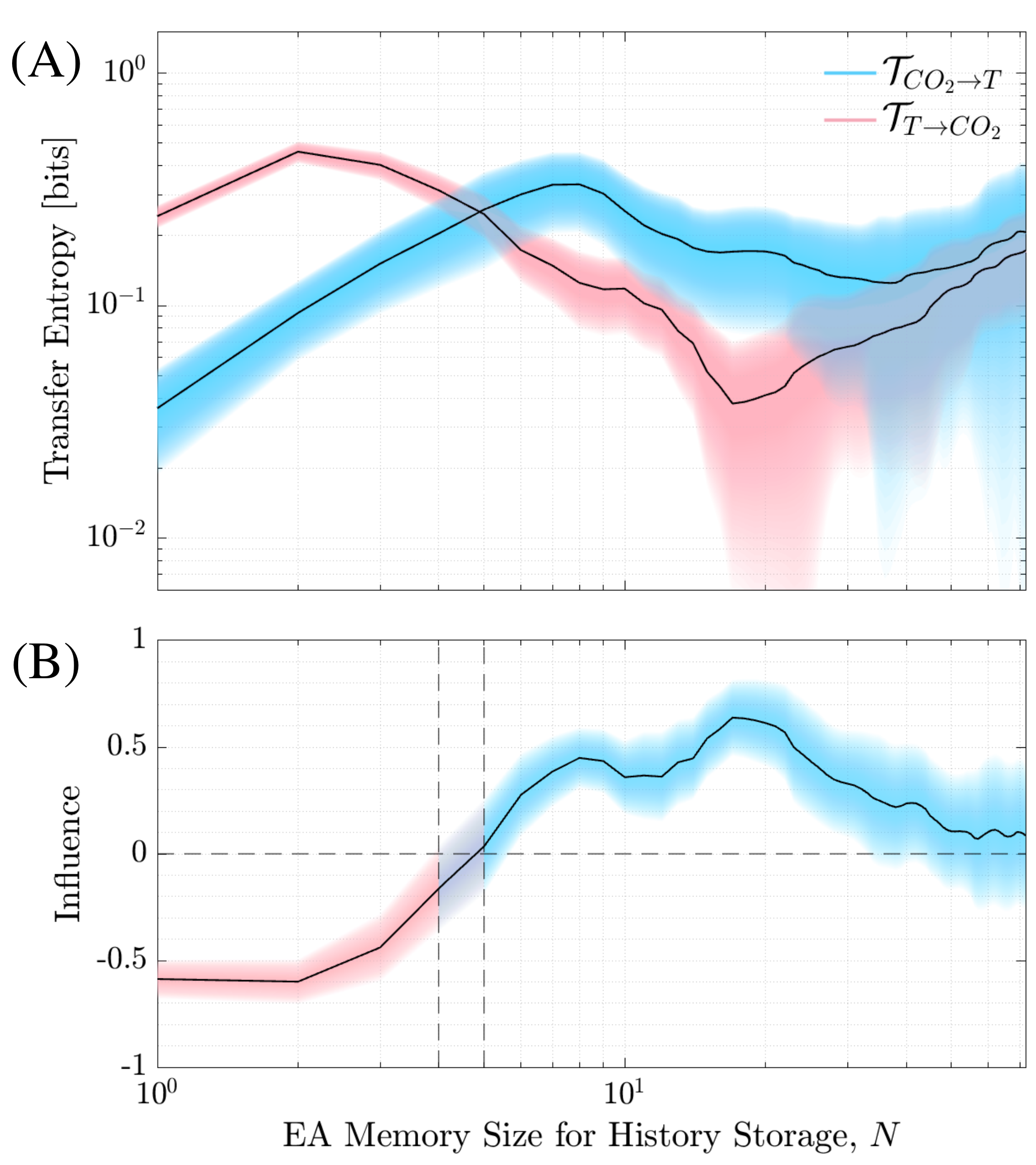}
    \caption{
    (A) Transfer entropies between CO$_2$ and temperature sensors at taken from Mauna Loa as a function of the memory size $N$ (measured in months), used to compute them.
    Error bars go out to 5$\sigma$ and are computed from a bootstrapped ensemble as explained in the text.
    (B) The influence between the two plotted on a $\log-\log$ scale.
    To admit negative values, only the magnitude is plotted while the color scheme signifies the sign (direction) of influence.
    Note that the heights on either side are comparable, signifying similar influence judgements by EAs with different architectures. 
    For nearly all members of the ensemble, vanishing influence typically occurred between $4-5$ months, so vertical bars have been added to emphasize this region.
    }
    \label{fig: arxiv5}
\end{figure}

These data consist of almost $800$ monthly measurements. 
We examine the transfer entropy between measured CO$_2$ and temperature processes in order to classify regimes of influence between them. 
Bootstrapping is used to get an ensemble of data streams in order to understand qualitatively the associated errors in the information measures. 
This is done by first choosing equal length substreams and then introducing Gaussian noise with standard deviation equal to the significant digit in the original data. 
We experimented with substreams of many different length and found the results robust for lengths up to $\sim20-30$ months, beyond which data volume became an issue as can be seen in the overlap of error bars in Fig. \ref{fig: arxiv5} (A).

Figure \ref{fig: arxiv3} shows that the influence an EA would ascribe between CO$_2$ and temperature does, indeed, depend on the memory usage. 
For $N t < 4$ months, a period associated with changing weather, an EA would judge temperature to influence CO$_2$. 
For $N t > 6$ months, a period associated with climate, an EA would judge the opposite influence to be true.
The two data streams yield opposite conclusions about which process influences the other dependent on the architecture of the EA.  
Thus, unless an agreement is reached about the proper memory usage $N$, no consensus will emerge in an ensemble of EAs examining the data streams.\\

\section{Discussion and Conclusion}\label{sec: 5}
Our results show that dealing with multiple epistemic agents is problematic on two fronts. 
Even if the EAs have access to identical data to inform their beliefs about histories, the differences in their temporal sampling and memory usage can result in a lack of consensus on influence judgements. 
This is under ideal conditions, where the processes are perfectly accessible to all the EAs. 
Moving away from these ideal conditions, we believe, can only make the problem worse. 
Our results may have implications for issues surrounding the difficulty of reaching consensus under conditions of ideological polarization.

These results have implications for model-agnostic machine learning. 
Deep Bayesian architectures utilize information theoretic measures to {\it learn} from data. 
However, such learning demands efficient storage of belief distributions in lieu of enormous datasets, and is therefore subject to historical sampling. 
We predict that algorithms with memory usage optimized to specific hardware architectures  will encounter the consensus problem we describe when compared with identical algorithms running on different architectures.

Finally, since our work reveals how problematic reaching consensus is for multiple epistemic agents under ideal conditions, it begs further questions concerning the judgements made by groups of thinking entities. 
To what extent do these results hold for more than two data/qualia streams? 
Transfer entropy has been shown to be able to detect information {\it circuits} in groups of interacting agents \cite{bettencourt2008identification}.  
Based on our results, one can investigate the influence structure of these circuits, and how spatial and temporal coarse graining affect judgements.
These questions are crucial in anticipation of extending this work to the dynamics of group formation in  models of interaction in social organisms \cite{jiang2017identifying, RVAHDATI2019105867}. 
\bibliographystyle{apsrev4-1}
\bibliography{main_ArXiV}

\begin{appendices}
\section{A1. Solving the Equations of Motion}

\subsection{Dimensionalization}

The equations of motion can be written as the vector process
\ee{
\frac{d\bs{X}}{dt}&=-\bs{A}\cdot\bs{X}+\bs{B}\cdot\bs{\eta}\nonumber\\
&\text{where  }\bs{A}=
\begin{bmatrix}
\alpha_1 & -\alpha_1\\
-\alpha_2 & \alpha_2
\end{bmatrix}\text{  and  }
\bs{B}=
\begin{bmatrix}
\beta_1 & 0\\
0 & \beta_2
\end{bmatrix}.\nonumber
}
and the white noise first and second moments read $\langle\bs{\eta}\rangle=\bs{0}$ and $\langle\bs{\eta}(t)\cdot\bs{\eta}(t')^T\rangle=\mathds{1}\delta(t-t')$.
We begin by scaling the variables using the length and time scales constructed in the main text,
\begin{align}
\bs{X}&\rightarrow \ell\tilde{\bs{X}}\hspace{10pt}t\rightarrow \tau \tilde{t} \hspace{10pt} \bs{\eta}\rightarrow\tau^{-\frac{1}{2}}\tilde{\bs{\eta}},
\end{align}
and introducing the asymmetry parameters $a,b\in[-1,1]$, defined as
\ee{
a = \frac{\alpha_1-\alpha_2}{\alpha_1+\alpha_2}\hspace{10pt} b=\frac{\beta_1-\beta_2}{\beta_1+\beta_2},
}
the scaled operators become
\ee{
\tilde{\bs{A}}=\frac{1}{2}
\begin{bmatrix}
1+a & -1-a\\
-1+a & 1-a
\end{bmatrix}\ \ \tilde{\bs{B}}=\frac{1}{2}
\begin{bmatrix}
1+b & 0\\
0& 1-b
\end{bmatrix}.\nonumber
}
The equations of motion are then the same as before
\begin{align*}
    \frac{d\tilde{\bs{X}}}{d\tilde{t}}=-\tilde{\bs{A}}\cdot\tilde{\bs{X}}+\tilde{\bs{B}}\cdot\tilde{\bs{\eta}}.
\end{align*}
Henceforth we drop the tildes and work in the dimensionless variables.

\subsection{Diagonalization}
In this section we will diagonalize the equations of motion in order to show that our system is Gaussian.
The eigenvalues of $\bs{A}$ are $0$ and $1$.
The first of these represents the translational symmetry of the system; $X_i\mapsto X_i+c$, for constant $c$, leaves the system invariant.
The corresponding eigenvectors are
\ee{
\bs{v}_0 = \frac{1}{\sqrt{2}}
\begin{bmatrix}
1\\
1
\end{bmatrix}
\hspace{10pt}
\bs{v}_1=\frac{1}{\sqrt{2(1+a^2)}}
\begin{bmatrix}
-(1+a)\\
1-a
\end{bmatrix}
\nonumber
}
From these we construct the similarity matrix that takes us into the diagonal frame, $\bs{S}=[\bs{v}_0 \bs{v}_1]$.
There the variables are the displacement and the center of mass, $\bs{Y}=\bs{S}^{-1}\cdot\bs{X}$,
\ee{
Y_2 = \sqrt{\frac{1+a^2}{2}}(X_2-X_1)\hspace{10pt}
Y_1 = \frac{1-a}{\sqrt{2}} X_1+\frac{1+a}{\sqrt{2}} X_2,\nonumber 
}
respectively.
Note that the center of mass is weighed by the deterministic coupling parameter.
Meanwhile, the rotated noise is $\bs{\xi}=\bs{S}^{-1}\cdot\bs{B}\cdot{\bs{\eta}}$.
The original noise was uncorrelated and Gaussian --- we see that the rotated noise will have non-vanishing cross correlations
\begin{align*}
\langle\bs{\xi}(t)\cdot\bs{\xi}^T(t')\rangle
&=\bs{S}^{-1}\cdot\bs{B}\cdot\langle\tilde{\bs{\eta}}(t)\cdot\tilde{\bs{\eta}}^T(t')\rangle\cdot\bs{B}^T\cdot\bs{S}^{-T}\\
&=\frac{1}{4}
\begin{bmatrix}
(1+a^2)(1+b^2) & \sqrt{1+a^2}(a-2b+ab^2)\\
\sqrt{1+a^2}(a-2b+ab^2) & 1+ a^2 -4 a b+b^2+ a^2b^2
\end{bmatrix}\delta(t-t').
\end{align*}
Diagonalizing the deterministic coupling of the degrees of freedom has introduced a stochastic coupling of the noise experienced by the degrees of freedom.
In this frame the equations of motion read
\begin{align*}
    \frac{dY_1}{dt}=-Y_1+\xi_1\hspace{10pt}
    \frac{dY_2}{dt}=\xi_2\hspace{10pt}\Rightarrow\hspace{10pt} \frac{d\bs{Y}}{d\tilde{t}}=-\bs{\Lambda}\cdot\bs{Y}+\bs{\xi},\nonumber
\end{align*}
where $\bs{\Lambda}=\text{diag}(1,0)$ is the eigenvalue matrix.
We recognize the first of these as a standard Wiener process, and the second as an Ornstein-Uhlenbeck process.
We say that they are uncoupled deterministicaly, but remain coupled stochastically due to the cross-correlation between noise. 
Both processes are Gaussian, so we see that our original system is Gaussian - it can be described completely by the covariance matrix since higher order correlations can all be derived from it.
\subsection{Solution}
The system can be solved by using an integration factor, $e^{\bs{\Lambda}t}$.
The solution in the diagonal frame is
\begin{align*}
    \bs{Y}(t)&=e^{-\bs{\Lambda}(t-t_0)}\cdot\bs{Y}(t_0)+\int_{t_0}^t\!ds\ \ e^{-\bs{\Lambda}(t-s)}\cdot\bs{\xi}(s)
\end{align*}
Note that since the eigenvalue matrix is idempotent, $\bs{\Lambda}^2=\bs{\Lambda}$, the matrix exponential can be reduced to 
\begin{align*}
    e^{\bs{\Lambda}t}&=\mathds{1}+\bs{\Lambda}t+\frac{1}{2}\bs{\Lambda}^2t^2+\frac{1}{3!}\bs{\Lambda}^3t^3\cdots\\
    &=\mathds{1}+\left(t+\frac{1}{2}t^2+\frac{1}{3!}t^3+\cdots\right)\bs{\Lambda}\\
    &=\mathds{1}+\left(e^t-1\right)\bs{\Lambda}\\
    &=
    \begin{bmatrix}
    \ e^{t} \ &\ 0 \ \ \\[0em]
    \ 0\ & \ 1 \ \ 
    \end{bmatrix}
\end{align*}
Rotating back to the original frame we have
\begin{align*}
    \tilde{\bs{X}}(t) = e^{-{\bs{A}}(t-t_0)}\cdot{\bs{X}}(t_0)+\int_{t_0}^t\! ds \ \ e^{-{\bs{A}}(t-s)}\cdot{\bs{B}}\cdot{\bs{\eta}}
\end{align*}
We could have gotten this directly from the original equations of motion, however we wanted to connect the system to well known Gaussian processes. 

\section{A2. Information Theory}
\subsection{Statistics}
Here we compute the mean and covariance of our vector process.
The mean is 
\ee{
{\bs{\mu}}(t)&=\langle{\bs{X}}(t)\rangle\nonumber\\ &=e^{-{\bs{A}}(t-t_0)}\cdot{\bs{X}}(t_0),\nonumber
}
from which one sees that the mean settles down to the perpendicular projection of the initial value of the process.
For completeness, in the original coordinate system
\begin{align}
    \mu_1(t)&=\frac{1}{2}(X_1(t_0)+X_2(t_0))+\frac{(1-e^{-(t-t_0)})a-e^{-(t-t_0)}}{2}(X_2(t_0)-X_1(t_0))\\
    \mu_2(t)&=\frac{1}{2}(X_1(t_0)+X_2(t_0))+\frac{(1-e^{-(t-t_0)})a+e^{-(t-t_0)}}{2}(X_2(t_0)-X_1(t_0)),
\end{align}

The covariance requires some work to find a decent form for
\begin{align*}
    \bs{\Sigma}(t.t')&=
    \langle\left(\tilde{\bs{X}}(t)-\langle\tilde{\bs{X}}(t)\rangle\right)\cdot\left(\tilde{\bs{X}}(t')-\langle\tilde{\bs{X}}(t')\rangle\right)^T\rangle\\
    &=\int_{t_0}^t\!ds \int_{t_0}^{t'}\! ds'\ \ e^{-\bs{A}(t-s)}\cdot\bs{B}\cdot\langle\bs{\eta}(s)\cdot\bs{\eta}(s')^T\rangle\cdot\bs{B}\cdot e^{-\bs{A}(t'-s')}\\
    &=\int_{t_0}^{\min(t,t')}\!ds \ \ e^{-\bs{A}(t-s)}\cdot\bs{B}\cdot\bs{B}\cdot e^{-\bs{A}(t'-s)}\\
    &=
\frac{(1+a^2)(1+b^2)-4ab}{8}
\begin{bmatrix}
1&1\\
1&1
\end{bmatrix}\min(t-t_0,t'-t_0)+\frac{a(1+b^2)-2b}{8}\nonumber\\
&\hspace{20pt}
\times\!\!
\begin{bmatrix}
(1\!+\!a)(1+e^{-|t-t'|}) & -(1\!-\!a)e^{-\max(t'\!-\!t,0)}\!+\!(1\!+\!a)e^{-\max(t\!-\!t',0)})\\
(1\!+\!a)e^{-\max(t'\!-\!t,0)}\!-\!(1\!-\!a)e^{-\max(t\!-\!t',0)} & -(1\!-\!a)(1+e^{-|t\!-\!t'|})
\end{bmatrix}\!\!
\nonumber\\
&\hspace{40pt}\times (1\!-\!e^{-\min(t\!-\!t_0,t'\!-\!t_0)})+\frac{1+b^2}{16}e^{-|t-t'|}
\begin{bmatrix}
(1+a)^2&-(1-a^2)\\
-(1-a^2) & (1-a)^2
\end{bmatrix}(1-e^{-2\min(t-t_0,t'-t_0)})
\end{align*}
Notice that the first term grows with $\min(t-t_0,t'-t_0)$, representing the diffusion of the underlying Wiener process- this dominates the covariance at large times. 
The second term is the slow contribution coming form the coupling, while the third term is the fast contribution - the latter grows at twice the rate of the former.
The parenthetical expressions to the right of both of these terms saturate to unity for long times, $t,t'\gg t_0$.
There is an interesting space of times early on when these terms are large enough compared to the first term that the covariance fluctuates.
Since $\bs{\Sigma}(t,t')^T=\bs{\Sigma}(t',t)$, without loss of generality we choose $t'=t+\Delta t>t$, which gives us a little space saving since we can replace
\begin{align*}
    \min(t-t_0,t'-t_0) = t-t_0\hspace{20pt}
    \max(t-t',0)=0\hspace{10pt}
    \max(t'-t,0)=\Delta t
\end{align*}
so that
\begin{align}
\bs{\Sigma}(t,t+\Delta t) &=
\frac{(1+a^2)(1+b^2)-4ab}{8}
\begin{bmatrix}
1&1\\
1&1
\end{bmatrix}
(t-t_0)\nonumber\\
&\hspace{20pt}
+\frac{a(1+b^2)-2b}{8}\!\!
\begin{bmatrix}
(1\!+\!a)(1+e^{-\Delta t}) & -(1\!-\!a)e^{-\Delta t}\!+\!(1\!+\!a)\\
(1\!+\!a)e^{-\Delta t}\!-\!(1\!-\!a) & -(1\!-\!a)(1+e^{-\Delta t})
\end{bmatrix}\!\! (1\!-\!e^{-(t\!-\!t_0)})
\nonumber\\
&\hspace{40pt}+\frac{1+b^2}{16}e^{-\Delta t}
\begin{bmatrix}
(1+a)^2&-(1-a^2)\\
-(1-a^2) & (1-a)^2
\end{bmatrix}(1-e^{-2(t-t_0)}).
\end{align}
This form shows us that the fast term decays with $\Delta t$, representing a short timescale correlation.
Meanwhile, the slow term gives a non-vanishing constant contribution to the long timescale correlations.
In either case, these are subdominant to the diffusion term. 

\subsection{Belief Distribution}
The EA's belief distribution over possible paths is required.
Since our process is Gaussian, the means and covariance between two histories are sufficient to completely describe the distribution.
We denote the former $\mu_i(t) = \langle X_i(t)\rangle$, and the latter with $\Sigma_{ij}(t,t') = \langle (X_i(t)-\langle X_i(t)\rangle)(X_j(t')-\langle X_j(t')\rangle\rangle$.
It is helpful to define the two instance covariance matrix via
\begin{align}
    \bs{\Sigma}(t,t') = 
    \begin{bmatrix}
        \Sigma_{11}(t,t) & \Sigma_{12}(t,t')\\
        \Sigma_{21}(t',t) & \Sigma_{22}(t',t')
    \end{bmatrix}.
\end{align}
Choosing a time discretization, $\Delta t$, we write the two histories $\bs{x} = [X_1(t),X_2(t),X_1(t-\Delta t),X_2(t-\Delta t),\dots,X_1(t-N\Delta t),X_2(t-N\Delta t)]$, the mean history $\bs{\mu}_N = [\mu_1(t),\mu_2(t),\mu_1(t-\Delta t),\mu_2(t-\Delta t),\dots,\mu_1(t-N\Delta t),\mu_2(t-N\Delta t)]$ and the covariance history
\begin{widetext}
\begin{align}
    \bs{\Sigma}_N(\Delta t)=
    \begin{bmatrix}
    \bs{\Sigma}(t,t) & \bs{\Sigma}(t,t-\Delta t) & \cdots & \bs{\Sigma}(t,t-N\Delta t)\\
    \bs{\Sigma}(t-\Delta t,t) & \bs{\Sigma}(t-\Delta t,t-\Delta t) & \cdots & \bs{\Sigma}(t-\Delta t,t-N\Delta t)\\
    \vdots &\vdots & \ddots & \ddots\\
    \bs{\Sigma}(t-N\Delta t,t) & \bs{\Sigma}(t-N\Delta t,t-\Delta t) & \cdots & \bs{\Sigma}(t-N\Delta t,t-N\Delta t)
    \end{bmatrix}
\end{align}
\end{widetext}
then the Gaussian approximation for the distribution over two histories reads
\begin{align}
    \rho(\bs{x};\bs{\mu}_N,\bs{\Sigma}_N) \!=\! \frac{1}{(2\pi)^N|\bs{\Sigma}_N|^\frac{1}{2}}\exp\!\!\left(\!\!-\frac{1}{2}(\bs{x}\!-\!\bs{\mu}_N)\Cdot\bs{\Sigma}_N^{-1}\Cdot(\bs{x}\!-\!\bs{\mu}_N)\!\!\right).
\end{align}
We mention that this form is amenable to numerical computations, since removing certain rows and columns gives the matrices necessary for the computation of the mutual information, as we shall now see.

\subsection{Information Measures}
With the belief distribution we can exactly compute the transfer entropy.
Consider subdividing our history vector into two disjoint parts, $\bs{x}=[\bs{x}_A \ \bs{x}_B]$.
Then the history mean breaks up similarly, while the history covariance breaks into 
\begin{align}
    \bs{\Sigma}=
    \begin{bmatrix}
    \bs{\Sigma}_{AA} & \bs{\Sigma}_{AB}\\
    \bs{\Sigma}_{BA} & \bs{\Sigma}_{BB}
    \end{bmatrix}.
\end{align}
With these, the mutual information between the disjoint subsets of events is the well known result for multivariate Gaussians,
\begin{align}\label{eq: MI of a Gaussian}
    M[A:B]=\frac{1}{2}\log\frac{|\bs{\Sigma}_{AA}||\bs{\Sigma}_{BB}|}{|\bs{\Sigma}|}.
\end{align}
It is clear that if the subsets are uncorrelated, $|\bs{\Sigma}|=|\bs{\Sigma}_{AA}||\bs{\Sigma}_{BB}|$, the logarithm is unity and the mutual information vanishes. 
Using Eq. \ref{eq: MI of a Gaussian} in the mutual information form of the transfer entropy, let's do the $N=1$ calculation explicitly. 
\begin{align*}
     \mathcal T_{2\rightarrow 1}\!&=\! M[X_1,t\!:\!X_1'\!\wedge\! X_2',t'=t-\Delta t]\!-\!M[X_1,t\!:\!X_1',t'=t-\Delta t]\\
     &=\frac{1}{2}\log{\frac{
     \begin{vmatrix}
     \Sigma_{11}(t,t)
     \end{vmatrix}
     \begin{vmatrix}
     \Sigma_{11}(t',t') & \Sigma_{12}(t',t')\\
     \Sigma_{21}(t',t') & \Sigma_{22}(t',t')
     \end{vmatrix}
     }{
     \begin{vmatrix}
     \Sigma_{11}(t,t) & \Sigma_{11}(t,t') & \Sigma_{12}(t,t')\\
     \Sigma_{11}(t,t') & \Sigma_{11}(t',t') & \Sigma_{12}(t',t')\\
     \Sigma_{12}(t,t') & \Sigma_{12}(t',t') & \Sigma_{22}(t',t')
     \end{vmatrix}
     }}-\frac{1}{2}\log{\frac{
     \begin{vmatrix}
     \Sigma_{11}(t,t)
     \end{vmatrix}
     \begin{vmatrix}
     \Sigma_{11}(t',t')
     \end{vmatrix}
     }{
     \begin{vmatrix}
     \Sigma_{11}(t,t) & \Sigma_{11}(t,t')\\
     \Sigma_{11}(t,t') & \Sigma_{11}(t',t')
     \end{vmatrix}}}
\end{align*}
One can play around with this expression, but, in the end, cannot escape taking the determinants and creating an algebraic mess of gargantuan size--- the expressions for $N>1$ are even more space consuming, since the determinants can gain up to $2$ additional rows and columns each time $N$ goes up by $1$.
Rather than opting for the analytical expressions, this is where we went the numerical route, implementing the determinants pointwise across a finite domain of all relevant variables.
This allowed us to compute values of $N$ up to about $100$, at which point computational time created a large bottleneck which prevented further investigation.
It is clear that the limits $N\rightarrow\infty,\Delta t\rightarrow 0$ are interesting since they correspond to ideal epistemic agents.
The transfer entropy in the opposite direction is computed by simply exchanging $1\leftrightarrow 2$.
With the two information flows, the influence is easily computed. 

\section{A3. Mauna Loa Data}
We acquired our data for atmospheric CO$_2$ content and temperature for the Mauna Loa observation site from the NOAA and WMO websites\cite{dataMaunaLoa, dataMaunaLoa2}.
The data spans the years $1958-2021$, with measurements taken at monthly intervals.
The CO$_2$ data is measured in parts per million, and has an accuracy of $10^{-2}$.
The temperature data represents the mean daily temperature in Celsius and has an accuracy of $10^{-1}$.
\subsection{Bootstrapping}
Note that $\Delta t=1$month is fixed by the data, so we explored the affect of memory size, $N$, on EA judgements.
To create a data ensemble that would allow us to measure the errors in influence we bootstrapped our data to create $1000$ smaller data sets.
Each smaller dataset was constructed as follows after fixing $N$.
A random {\it present} time, $t$, is chosen uniformly from the data between $2004-2021$.
The original data is taken from $1958-t$, and arithmetic noise is added at each month drawn from a Gaussian distribution with $0$ mean and standard deviation equal to the accuracy of the data. 
Transfer entropy and influence are computed for this noised subset of data for values of $N=1,2,\dots 100$, corresponding to EA memory capacities of up to $N\Delta t\sim8$years.
We then used a peaked smoothing kernel, $[0.09, 0.18, 0.46, 0.18, 0.09]$, to clean up the simulation results. 
This is repeated $1000$ times to generate the full ensemble.
For fixed $N$, statistics are run on the ensemble, with means and standard deviations computed via unbiased estimators.
\end{appendices}

\end{document}